\documentclass[showpacs,preprintnumbers,amsmath,amssymb]{revtex4}

\usepackage{amsmath,amssymb,amsthm,epsf,epsfig}
\usepackage{amsmath,amssymb}
\usepackage{graphicx}
\usepackage{dcolumn}
\usepackage{bm}

\begin{document}

\title{Position Dependent Mass Schr\"{o}dinger Equation and
Isospectral Potentials : Intertwining Operator approach}

\author{Bikashkali Midya}\email{bikash.midya @ gmail.com}
\author{B. Roy}\email{barnana @ isical.ac.in}
 \author{R. Roychoudhury} \email{raj @ isical.ac.in}
\affiliation{Physics \& Applied Mathematics Unit\\ Indian
Statistical Institute \\ Kolkata 700108\\ India }


\begin{abstract} Here we have studied first and second-order
intertwining approach to generate isospectral partner potentials
of position-dependent (effective) mass Schr\"{o}dinger equation.
The second-order intertwiner is constructed directly by taking it
as second order linear differential operator with position
depndent coefficients and the system of equations arising from the
intertwining relationship is solved for the coefficients by taking
an ansatz. A complete scheme for obtaining general solution is
obtained which is valid for any arbitrary potential and mass
function. The proposed technique allows us to generate isospectral
potentials with the following spectral modifications: (i) to add
new bound state(s), (ii) to remove bound state(s) and (iii) to
leave the spectrum unaffected. To explain our findings with the
help of an illustration, we have used point canonical
transformation (PCT) to obtain the general solution of the
position dependent mass Schrodinger equation corresponding to a
potential and mass function. It is shown that our results are
consistent with the formulation of type A ${\cal{N}}$-fold
supersymmetry \cite{r27,r28} for the particular case ${\cal{N}}=1$
and ${\cal{N}}=2$ respectively.
\end{abstract}
\maketitle

\vspace{.2 cm}
\section{\label{}Introduction}
There is a growing interest nowadays to design systems whose
Hamiltonians have given spectral characteristics. In this context,
the idea of designing potentials with prescribed energy spectra is
worth investigating. Some progress in this area has been done by
restricting the construction of potentials isospectral to a given
initial one except a few energy values through the usage of
Darboux transformation \cite{r1}, factorization method \cite{r2},
supersymmetric quantum mechanics (SUSYQM) \cite{r3,r23} and other
related techniques. The underlying idea of most of these
procedures has been summarized in an algebraic scheme known as
intertwining approach. In general, the objective of the
intertwining is to construct the so-called intertwining operator
$\mathcal{L}$ which performs an intertwining between an initial
solvable Hamiltonian $\mathcal {H}$ and a new solvable one
$\mathcal{\bar{H}}$ with slightly modified spectrum such that
\begin{equation}\mathcal {L} \mathcal {H} = \bar {\mathcal {H}} \mathcal
{L},~~~~~\bar{\psi}(x) = \mathcal{L} \psi(x)\label{e60}
\end{equation}
 The ingredients to implement the intertwining are seed solutions of the initial stationary
 Schr\"odinger equation associated to factorization energies less than or equal to the ground state
  energy of $\mathcal {H}$. If $\mathcal{L}$ is a first order differential operator, the
standard SUSYQM, with supercharges built of first order Darboux
transformation operators, and the factorization method are
recovered. On the other hand, if higher order differential
operators are involved in the construction of $\mathcal {L}$, it
gives rise to higher order SUSYQM \cite{r100}. It is possible to generate
families of isospectral Hamiltonians by either of the two ways:\\
(i) by iteration of first order Darboux transformations. Every
chain of ${\cal{N}}$ first order Darboux transformation creates a
chain of exactly solvable Hamiltonians
$\mathcal{H}_0,\mathcal{H}_1,\mathcal{H}_2...\mathcal{H}_{\cal{N}}$
\cite{r40}. Hence the intertwining operator
$\mathcal{L}^{({\cal{N}})}$ between the initial Hamiltonian
$\mathcal{H}_0$ and the final Hamiltonian $\mathcal{H}_{\cal{N}}$
can always be presented as a product of ${\cal{N}}$ first order
Darboux transformation operators between every two juxtaposed
Hamiltonians $\mathcal{H}_0
,\mathcal{H}_1,...,\mathcal{H}_{\cal{N}}$,
$$\mathcal{L}^{({\cal{N}})} = \mathcal{L}_{\cal{N}}
\mathcal{L}_{{\cal{N}}-1}...\mathcal{L}_2\mathcal{L}_1,~~~
\mathcal{H}_p\mathcal{L}_p = \mathcal{L}_p\mathcal{H}_{p-1},~~ p =
1,2,...,{\cal{N}}$$ (ii) by looking for the ${\cal{N}}$-th order
intertwining operator directly, expressing the intertwiner as a
sum of ${\cal{N}}+1$ terms $g_i(x) \frac{d^i}{dx^i}, i =
0,1,\cdots {\cal{N}}$, and solving the system of equations
resulting from the intertwining relationship for the $g_i(x)$'s.

At this point it is appropriate to mention that the quantum
mechanical systems with position dependent (effective) mass
\cite{r4} have attracted a lot of interest due to their relevance
in describing the physics of many microstructures of current
interest such as semiconductor heterostructures \cite{r5}, quantum
dots \cite{r6}, helium clusters and metal crystals \cite{r8} etc.
Recently, the intertwining operator method has been applied to
Schr\"odinger equation with position dependent (effective) mass to
construct first-order and chains(iterations) of first-order
Darboux transformations and the connection between the first-order
Darboux transformation and effective mass supersymmetry
(factorization) was shown \cite{r10}. Subsequently Darboux
transformations of arbitrary order for position dependent mass
Schr\"odinger equation was derived and factorization of the n-th
order transformation into first order transformations and
existence of a reality condition for the transformed potential was
shown \cite{r12}. In the standard supersymmetric (SUSY) approach
for effective mass Hamiltonians \cite{r7}, the ladder operators
are taken as first order differential operators similar to
constant mass case, but now they depend on both superpotential and
mass function. As a result one obtains two partner potentials with
the same effective mass sharing identical spectra upto the zero
mode of the supercharge. Second order supersymmetric approach
(2-SUSY)was used in ref \cite{r26} for describing dynamics of a
quantum particle with a position dependent mass. A compact
expression for 2-SUSY isospectral pairs was derived in terms of
senond order superpotential and the mass function. A detailed
analysis has been given about zero mode equations of second order
supercharges and possible reduction of 2-SUSY scheme to first
order SUSY. Recently, a generalization of standard SUSY known as
higher derivative SUSY or ${\cal{N}}$-fold SUSY \cite{r27} was given for
position dependent mass Hamiltonians. This method keeps the basic
superalgebra intact but differs from the standard first order SUSY
in that the supercharges are represented  as $\cal{N}$-th order $({\cal{N}}>1)$
differential operators.

In this article an attempt is made to generate isospectral
potentials of position dependent mass Schr\"odinger equation
(PDMSE) by applying first and second order intertwining technique.
Specifically, the second order intertwiner is constructed by
taking it as

$$\mathcal {L} = \frac{1}{M(x)}\frac{d^2}{dx^2} + \eta (x) \frac{d}{dx} + \gamma (x)$$

where $M(x)$ is the mass function and $\eta(x)$, $\gamma(x)$ are
to be determined. Substituting this in the intertwining
relationship (\ref{e60}), we have been able to solve the
apparently intricate system of equations for $\eta(x)$ and
$\gamma(x)$ by assuming an ansatz. As mentioned earlier, the
closed form formulas for ${\cal{N}}$-th order intertwining
operators and a pair of isospectral Hamiltonians with position
dependent mass was already reported by Tanaka \cite{r27}, for an
arbitrary value of ${\cal{N}}$ without recourse to any ansatz.

The motivation for constructing second order intertwining operator
directly comes from the observation that although an
${\cal{N}}$-th order intertwining operator can be expressed
formally as a product of ${\cal{N}}$ first order intertwining
operators, it does not necessarily mean that a system constructed
by an ${\cal{N}}$-th order intertwining operator is equivalent to
one constructed by ${\cal{N}}$ successive applications of each
$1$st order operator. In fact, it was shown in ref \cite{r29}, by
comparing the two approaches in the constant mass scenario, that
the former is more general than the latter. The advantage of the
direct method used here over the iterative method is that one can
generate second-order isospectral partner potentials directly from
the initial potential i.e. one need not go through the first-order
intertwining technique. We shall see that as in the case of
constant mass scenario, it is possible to generate isospectral
potentials with some spectral modification: (i) to add new bound
state(s) (ii) to remove bound state(s) and (iii) to leave the
spectrum unaffected. In this context a natural question is: What
is the utility of finding the isospectral potentials in the
position dependent mass background? To answer this let us note
that in different areas of possible applications of low
dimensional structures as already mentioned, there is need to have
energy spectrum which is predetermined. For example, in the
quantum well profile optimization, isospectral potentials (by
deleting or creating bound states at a particular energy of the
original potential) are generated through supersymmetric quantum
mechanics. This is necessary because a particular effect (such as
intersubband optical transitions in a quantum well) may be grossly
enhanced by achieving the resonance conditions e.g. appropriate
spacings between the most relevant states and also by tailoring
the wave functions so that the (combinations of) matrix elements
relevant for this particular effect are maximized \cite{r30}. This
is particularly important for studying higher order nonlinear
processes.

 The organization of the paper is as follows: in section II and III
we have explained the first and second-order intertwining
techniques respectively with possible spectral modifications, with
the help of a suitable example given in the Appendix. Also, the
connection of our approach to type A ${\cal{N}}$-fold SUSY is shown in these sections.
Section IV is kept for discussions and comments.
\section{\label{} First order Intertwining} We consider the
following two one-dimensional effective mass Schrodinger
Hamiltonians (Bendaniel-Duke form) \cite{r21} with the same
spectrum but with different potential
\begin{equation}
\mathcal{H}\psi=E\psi~,~~\mathcal{H}=-\left[\frac{d}{dx}\left(\frac{1}{M(x)}\right)\frac{d}{dx}\right]+V(x)\label{e15}
\end{equation}
and
\begin{equation}
    \bar{\mathcal{H}}\bar{\psi}=E\bar{\psi}~,~~\bar{\mathcal{H}}
    =-\left[\frac{d}{dx}\left(\frac{1}{M(x)}\right)\frac{d}{dx}\right]+\bar{V}(x)\label{e16}
\end{equation}
We connect the Hamiltonians (\ref{e15}) and (\ref{e16}) by means
of the intertwining technique. To this end,
 we look for an operator $\mathcal {L}$ that satisfies the
 relation (\ref{e60}). Without loss of generality let us consider the first order
intertwining operator \cite{r10}, as
\begin{equation}
\mathcal{L}=\frac{1}{\sqrt{M(x)}}\frac{d}{dx}+A(x)\label{e17}
\end{equation}
 Now using the intertwining relation (\ref{e60}) and equating the coefficients of
like order of derivatives we obtain
\begin{equation}
\bar{V}=V+\frac{2A'}{\sqrt{M}}-\frac{3M'^2}{4M^3}+\frac{M''}{2M^2}\label{e29}
\end{equation}
and
\begin{equation}
A(\bar{V}-V)=-\frac{A'M'}{M^2}+\frac{V'}{\sqrt{M}}+\frac{A''}{M}\label{e30}
\end{equation}
where `prime' denotes differentiation with respect to $x$.\\
 Now using
(\ref{e29}) , equation (\ref{e30}) reduces to
\begin{equation}
\frac{A''}{\sqrt{M}}-\frac{A'M'}{M^{\frac{3}{2}}}-\frac{AM''}{2M^{\frac{3}{2}}}+\frac{3AM'^{2}}{4M^{\frac{5}{2}}}
-2AA'+V'=0\label{e36}
\end{equation}
Integrating equation (\ref{e36}) we get
\begin{equation}
\frac{A'}{\sqrt{M}}-\frac{AM'}{2M^{\frac{3}{2}}}-A^2+V=\mu\label{e31}
\end{equation}
where $\mu$ is a constant of integration. Now we substitute
\begin{equation}
A(x)=-\frac{K}{\sqrt{M}}\label{e57}
\end{equation}
where K=K(x) being an auxiliary function, in equation (\ref{e31})
we obtain the following Riccati equation
\begin{equation}
-\frac{K'}{M}+\frac{KM'}{M^2}-\frac{K^2}{M}+V=\mu\label{e58}
\end{equation}
The equation (\ref{e58}) can be linearized by the substitution
$K(x)=\frac{\mathcal{U'}(x)}{\mathcal{U}(x)}$. Substituting this
value of $K$ in equations (\ref{e57}) and (\ref{e58}) we get
\begin{equation}
A(x)=-\frac{\mathcal{U'}}{\sqrt{M}\mathcal{U}}\label{e59}
\end{equation}
and
\begin{equation}
-\frac{1}{M}\mathcal{U}''-\left(\frac{1}{M}\right)'\mathcal{U}'+V\mathcal{U}=\mu
\mathcal{U}\label{e18}
\end{equation}
respectively. The equation (\ref{e18}) is similar to equation
(\ref{e15}) with $E=\mu$, $\mu$ is sometimes called factorization
energy and $\mathcal{U}(x)$ is called seed solution. It should be
noted here that $\mathcal{U}(x)$ need not be normalizable solution
of (\ref{e18}). However, for $A(x)$ to be well defined (without
singularity), $\mathcal{U}$ must not have any zeroes on the real
line. For this, we shall restrict $\mu\leq E_0$ throughout this
article, $E_0$ being the ground state energy eigenvalues of the
equation (\ref{e15}). Once we can determine the solution
$\mathcal{U}$ of (\ref{e18}) then we shall able to construct the
intertwiner $\mathcal{L}$, the isospectral partner $\bar{V}$ and
its bound state eigenvalues $\bar{\psi}(x)$ with the help of
following relations
\begin{equation}
\bar{V}=V-\frac{2\mathcal{U''}}{M\mathcal{U}}+\frac{2\mathcal{U'}^2}{M\mathcal{U}^2}+\frac{M'\mathcal{U'}}
{M^2\mathcal{U}}+\frac{M''}{2M^2}-\frac{3M'^2}{4M^3}\label{e70}
\end{equation}
\begin{equation}
\hspace{-3.8
cm}\mathcal{L}=\frac{1}{\sqrt{M}}\left(\frac{d}{dx}-\frac{\mathcal{U'}}{\mathcal{U}}\right)\label{e71}
\end{equation}
and
\begin{equation}
\hspace{-2 cm}\bar{\psi}(x)\propto
\mathcal{L}\psi=\frac{1}{\sqrt{M}} \left[\frac{d}{dx}-(ln~
\mathcal{U})'\right]\psi\label{e21}
\end{equation}
The intertwiner $\mathcal {L}$ cannot be used to generate wave
function of $\bar {\mathcal {H}} $ at the factorization energy
$\mu$, because $\mathcal{LU}=0$. We are showing below with the
help of supersymmetry how $\bar {\psi_{\mu}}$  can be obtained
from the relation $\mathcal{L}^{\dagger}\bar{\psi_{\mu}} = 0$,
$\mathcal{L}^{\dagger}$ being the adjoint of $\mathcal{L}$ and is
given by
\begin{equation}
\mathcal{L}^\dagger=\frac{1}{\sqrt{M}}
\left(-\frac{d}{dx}-\frac{\mathcal{U'}}{\mathcal{U}}+\frac{M'}{2M}\right)\label{e63}
\end{equation}
For this we calculate $\mathcal{L}^\dagger \mathcal{L}$ and
$\mathcal{L}\mathcal{L}^\dagger$  given by
\begin{equation}
\hspace{-2.4 cm}\mathcal{L}^\dagger
\mathcal{L}=-\frac{1}{M}\frac{d^2}{dx^2}+\frac{M'}{M^2}\frac{d}{dx}+\left(\frac{\mathcal{U''}}{M\mathcal{U}}
-\frac{M'\mathcal{U'}}{M^2\mathcal{U}}\right)\label{e64}
\end{equation}
and
\begin{equation}
\mathcal{L}
\mathcal{L}^\dagger=-\frac{1}{M}\frac{d^2}{dx^2}+\frac{M'}{M^2}\frac{d}{dx}+\left(\frac{2\mathcal{U'}^2}
{M\mathcal{U}^2}-\frac{\mathcal{U''}}{M\mathcal{U}}
+\frac{M''}{2M^2}-\frac{3M'^2}{4M^3}\right)\label{e65}
\end{equation}
Now from the equation (\ref{e18}) we have
\begin{equation}
V=\frac{\mathcal{U''}}{M\mathcal{U}}
-\frac{M'\mathcal{U'}}{M^2\mathcal{U}}+\mu\label{e66}
\end{equation}
Substituting this value of $V$ in (\ref{e70}) we obtain
\begin{equation}
\bar{V}=-\frac{\mathcal{U''}}{M\mathcal{U}}
+\frac{2\mathcal{U'}^2}{M\mathcal{U}^2}+\frac{M''}{2M^2}-\frac{3M'^2}{4M^3}+\mu\label{e67}
\end{equation}
Now using (\ref{e66}) and (\ref{e67}) in (\ref{e64}) and
(\ref{e65}) we get
\begin{equation}
\mathcal{L}^\dagger
\mathcal{L}=-\frac{1}{M}\frac{d^2}{dx^2}+\frac{M'}{M^2}\frac{d}{dx}+V-\mu=\mathcal{H}-\mu\label{e68}
\end{equation}
and
\begin{equation}
\mathcal{L}\mathcal{L}^\dagger=-\frac{1}{M}\frac{d^2}{dx^2}+\frac{M'}{M^2}\frac{d}{dx}+\bar{V}-\mu=
\bar{\mathcal{H}}-\mu\label{e69}
\end{equation}
respectively. It is clear from the equation (\ref{e69}) that the
wave function of $\mathcal{\bar{H}}$ at the factorization energy
$\mu$ can be obtained by $\mathcal{L}^\dagger\bar{\psi}_{\mu}=0$
i.e.,
\begin{equation}
\bar{\psi}_{\mu}\propto exp\left[\int\left(
-\frac{\mathcal{U'}}{\mathcal{U}}+\frac{M'}{2M}\right)dx\right]=\frac{\sqrt{M}}{\mathcal{U}}\label{e72}
\end{equation}
It is to be noted that if $\mathcal{U}$ corresponds to the bound
state of $\mathcal{H}$, the wave function $\bar{\psi}_{\mu}(x)$
defined in (\ref{e72})is not normalized so that $\mu$ does not
belong to the bound state spectrum of $\mathcal{\bar{H}}$. If
$\mathcal{U}$ corresponds to the ground state wavefunction of
$\mathcal{H}$ then then the potential $\bar{V}$ has no new
singularity, except the singularity due to $V$, provided $M$ is
not singular and $M\neq 0.$ However, if we consider $\mathcal{U}$
to an arbitrary state other than ground state of $\mathcal{H}$
then $\bar{V}$ might contain extra singularities, which are not
present in $V$. If $\mathcal{U}$ is nodeless and unbounded at the
both end points then $\bar{\psi}_{\mu}(x)$ defined in (\ref{e72})
is normalizable, so that $\mu$ can be included
 in the bound state spectrum of $\mathcal{H}$ to generate $\bar{V}.$ In this case maximal set
of bound state wavefunctions of $\mathcal{\bar{H}}$ are given by
$\{\bar{\psi}_{\mu}, \mathcal{L}\psi\}$.\\
\subsection{First-order intertwining and type A 1-fold SUSY}
To show that the results obtained in the previous section are
consistent with the results of type A ${\cal{N}}$-fold SUSY, we
are going to mention the brief results of  type A $\cal{N}$-fold
SUSY formalism ( for details see \cite{r27} and references there).
Type A $\cal{N}$-fold SUSY is characterized by the type A monomial
space
\begin{equation}
\bar{\nu}_{\cal{N}}=\langle1,z,...z^{{\cal{N}}-1}\rangle\label{e99}
\end{equation}
preserved by $\tilde{\cal{H}}_{\cal{N}}:$
\begin{equation}
\tilde{\cal{H}}^-_{\cal{N}}=-A(z)\frac{d^2}{dz^2}-B(z)\frac{d}{dz}-C(z)\label{e100}
\end{equation}
where
\begin{equation}\begin{array}{llll}
A(z)= a_4 z^4+a_3z^3+a_2z^2+a_1z+a_0,~~{\cal{N}}\geq 3\\
B(z)=Q(z)-\frac{{\cal{N}}-2}{2} A'(z)\\
C(z)= \frac{({\cal{N}}-1)({\cal{N}}-2)}{12}
A''(z)-\frac{({\cal{N}}-1)}{2} Q'(z)+R\\
Q(z)= b_2 z^2+b_1z+b_0,~~~~~{\cal{N}}\geq 2\\
\end{array}\label{e101}
\end{equation}
$R$, $a_i, b_i$ are being constants. Applying the algorithm for
constructing type A $\cal{N}$-fold SUSY in PDM system \cite{r27},
one can construct the most general form of type A $\cal{N}$-fold
SUSY PDM quantum systems $(\cal{H},\bar{\cal{H}}, \cal{L})$ or
equivalently $(\cal{H}_{N}^{+},\cal{H}_{N}^{-}, P_{\cal{N}}):$
\begin{equation}
{\cal{H}}^{\pm}_{{\cal{N}}}=-\frac{1}{M} \frac{d^2}{dx^2} +
\frac{M'}{M^2}\frac{d}{dx} + V^{\pm} (x)\label{e102}
\end{equation}
\begin{equation}
P_{\cal{N}} = M(x)^{-\frac{\cal{N}}{2}} \prod_{k=0}^{{\cal{N}}-1}\left[\frac{d}{dx}+W(x)-\frac{{\cal{N}}M'(x)}{4M(x)}+\frac{{\cal{N}}-1-2k}{2}\frac{z''(x)}{z'(x)}\right]\label{e103}
\end{equation}
where
\begin{equation}
V_{+}= V^{-}+ 2{\cal{N}}\left(\frac{W'(x)}{M(x)}-\frac{M'(x)W(x)}{2M(x)^2}\right)\label{e104}
\end{equation}
\begin{equation}
W(x)=
\frac{d{\cal{W}}_{\cal{N}}^-(x)}{dx}-\frac{z''(x)}{z'(x)}+\frac{M'(x)}{2M(x)}
\end{equation}
\begin{equation}
\frac{d{\cal{W}}_{\cal{N}}^-(x)}{dx}=\frac{z''(x)}{2z'(x)}-\frac{M(x)B(z)}{2z'(x)}-\frac{M'(x)}{2M(x)}\label{e105}
\end{equation}
\begin{equation}
z'(x)^2=M(x) A(z)\label{e106}
\end{equation}
and the product of operators are ordered as
$$\prod_{k=0}^{{\cal{N}}-1} F_k =
F_{{\cal{N}}-1}F_{{\cal{N}}-2}...F_0$$ The solution space of the
type A Hamiltonians ${\cal{H}}^{\pm}_{\cal{N}}$ are given by
\begin{equation}
\nu^{\pm}_{\cal{N}} = e^{-{\cal{W}}^{\pm}_{\cal{N}}}
\langle1,z,...,z^{\cal{N}}\rangle|_{z=z(x)},~~{\mbox{where}}~~
{\cal{W}}^+_{\cal{N}}= - {\cal{W}}^-_{\cal{N}} +({\cal{N}}-1)~ ln
|z'(x)| - \frac{\cal{N}}{2}~ ln|M(x)|
\end{equation}
It is easily seen that $(\bar{V}-V)$ obtained in (\ref{e70})
coincides with $(V^+-V^-)$ in equation (\ref{e104}) (with
${\cal{N}}=1$) if one takes (comparing ${\cal{L}}$ with $P_1$)
\begin{equation}
\frac{d}{dx}~ln~ {\cal{U}}(x) = -W(x)+\frac{M'(x)}{4M(x)}\label{e107}
\end{equation}

\subsection{Example of first-order intertwining}
It may be emphasized that the results mentioned in section II are
most general and valid for any potential $V(x)$. However to
illustrate the above procedure with the help of an example we
shall need non-normalizable solutions of (\ref{e18}) corresponding
to a particular mass function $M(x)$. In Appendix A we have used
point canonical transformation approach(PCT) to solve the equation
(\ref{e18}). Here we are going to construct the isospectral
partners of the following potential obtained in Appendix A (we
have considered $p=\lambda=1$ for simplicity)
\begin{equation}
V(x)= \frac{[(a+b-c)^2-1]}{4} e^{x}+\frac{c(c-2)}{4} e^{-
x}\label{e10}
\end{equation}
 corresponding to the mass function
 \begin{equation}
 M(x)=\frac{1}{4}sech^2\left(\frac{ 1}{2}x\right)\label{e92}
 \end{equation}
 The bound state solutions and eigenstates of the
equation (\ref{e15}) are given by (see Appendix A)
\begin{equation}
 \psi_n(x)=\left(\frac{ (2n+\sigma+\delta+1) n!~\Gamma(n+\sigma+\delta+1)}
 {\Gamma(n+\sigma+1)\Gamma(n+\delta+1)}\right)^{1/2} \frac{e^{\frac{(\sigma+1)}{2}}x}{\left(1+e^{
 x}\right)^{\frac{\sigma+\delta+2}{2}}}\mathcal{P}_n^{(\sigma,\delta)}
 \left(\frac{1-e^{x}}{1+e^{x}}\right)\label{e14}
 \end{equation}
 and
 \begin{equation}
 E_n=n^2+n(\sigma+\delta+1)+\frac{(\sigma+1)(\delta+1)}{2}~,~~~n=0,1,2,...\label{e26}
 \end{equation}
 respectively, where $b=1-a+\sigma+\delta,~c=1+\sigma$ with $c>\frac{1}{2}~\mbox{and}~a+b-c+\frac{1}{2}>0.$
The seed solution $\mathcal{U}(x)$ and factorization energy $\mu$
are given by
\begin{equation}
 \mathcal{U}(x)=\alpha\frac{e^{\frac{c}{2}x}}{(1+e^{
x})^{\frac{a+b+1}{2}}}~ _2F_1\left(a,b,c,\frac{e^{ x}}{1+e^{
x}}\right) +\beta\frac{e^{\left(1-\frac{c}{2}\right) x}}{(1+e^{
x})^{\frac{a+b-2c+3}{2}}}~
 _2F_1\left(a-c+1,b-c+1,2-c,\frac{e^{ x}}{1+e^{
 x}}\right)\label{e13}
 \end{equation}
\begin{equation}
 \mu=-ab+\frac{(a+b+1)c}{2}-\frac{c^2}{2}\label{e9}
\end{equation}
respectively. The asymptotic behavior of the solution $\mathcal
{U}(x)$ given in (\ref{e13}), at both end points $\pm \infty$ are
given by \cite{r16}
\begin{equation}
 \mathcal{U}(x)\sim (A_1\alpha+B_1\beta) e^{-\frac{a+b-c+1}{2}x}+(A_2\alpha+B_2
 \beta)e^{-\frac{c-a-b+1}{2}x}~~~\mbox{as}
 ~x\rightarrow\infty\label{e22}
 \end{equation}
 where $$A_1=\frac{\Gamma(c)\Gamma(c-a-b)}{\Gamma(c-b)\Gamma(c-a)}~,
 ~~B_1=\frac{\Gamma(2-c)\Gamma(c-a-b)}{\Gamma(1-a)\Gamma(1-b)}$$
 $$A_2=\frac{\Gamma(c)\Gamma(a+b-c)}{\Gamma(a)\Gamma(b)}~,
 ~~B_2=\frac{\Gamma(2-c)\Gamma(a+b-c)}{\Gamma(a-c+1)\Gamma(b-c+1)}$$
 and
 \begin{equation}
\mathcal{U}(x)\sim \alpha e^{\frac{c}{2}x}+ \beta~
e^{(1-\frac{c}{2})x}~~~\mbox{as}
 ~x\rightarrow -\infty\label{e23}
 \end{equation}
 From these asymptotic behaviors it is clear that $\mathcal {U}(x)$ will unbounded
 at $x \rightarrow \infty$ if $|a+b-c|>1$ and it is unbounded at $x \rightarrow -\infty$ if $c<0 ~\rm or~ c>2$.
Therefore $\mathcal{U}(x)$ will nodeless at the finite part of the
$x$ axis if  $A_1\alpha+B_1\beta$, $A_2\alpha+B_2
 \beta$,$\alpha$ and $\beta$ are all positive and $|a+b-c|>1$, $c<0 ~\rm or~
 c>2$.\\Now we are going to generate isospectral potentials of the
 potential (\ref{e10}) with various possible spectral
 modifications.

 {\bf Deletion of the initial ground state :}
In this case the  factorization energy $\mu$ is equal to the
ground state energy $E_0$ giving $ a~ \rm {and/or}~ b = 0$ and
$\mathcal {U}(x)$ becomes the ground state wavefunction
$\psi_0(x)$ which is obtained from (\ref{e14}) as
 \begin{equation}
\mathcal{U}(x)=\psi_0(x)\propto
\frac{e^{\frac{c}{2}x}}{(1+e^x)^{\frac{a+b+1}{2}}}\label{e93}
\end{equation}
The isospectral partner of $V(x)$ given in equation (\ref{e10}),
is obtained using equation (\ref{e70}), (\ref{e10}), (\ref{e92})
and (\ref{e93}) and is given by
\begin{equation}
\bar{V}(x)=\frac{c^2-1}{4}e^{-x}+\frac{(a+b-c)(2+a+b-c)}{4}e^x+\frac{a+b}{2}
\label{e28}
\end{equation}
The above potential (\ref{e28}) can also be obtained from the
initial potential (\ref{e10}) by making the changes $a\rightarrow
a+1, b\rightarrow b+1, c\rightarrow c+1$, this property is known
as shape invariance \cite{r23}. Since $\psi_0(x)$ is bounded
solution $\bar{\psi}_{\mu}(x)=\frac{\sqrt{M}}{\psi_0}$ is
unbounded at $x\rightarrow\pm \infty$, so we have deleted the
ground state energy of $\mathcal {H}$ to obtain $\bar {V}(x)$.
 Therefore the eigenvalues of $\bar {H}$ are given by
 \begin{equation}
 \bar{E}_n=E_{n+1}=(n+1)^2+(n+1)(a+b)
 +\frac{c(a+b-c+1)}{2}~,~~~n=0,1,2...
 \end{equation} Corresponding bound
 state  wavefunctions of $\bar{V}(x)$ are obtained using equation (\ref{e21}) as
\begin{equation}
\bar{\psi}_n(x)\propto
\frac{e^{(1+\frac{c}{2})x}}{(1+e^x)^{(\frac{a+b+5}{2})}sech(\frac{x}{2})}
\mathcal{P}_{n}^{\left(c,a+b-c+1\right)}\left(-tanh\frac{x}{2}\right)~,~~
n=0,1,2...
\end{equation}
We have plotted the potentials $V(x)$ given in (\ref{e10}) and
$\bar {V}(x)$ given in (\ref{e28}) for
$a=5,b=0,c=3,\alpha=1,\beta=0$ in figure 1.
\begin{figure}[h]
\epsfxsize=3 in \epsfysize=2 in \centerline{\epsfbox{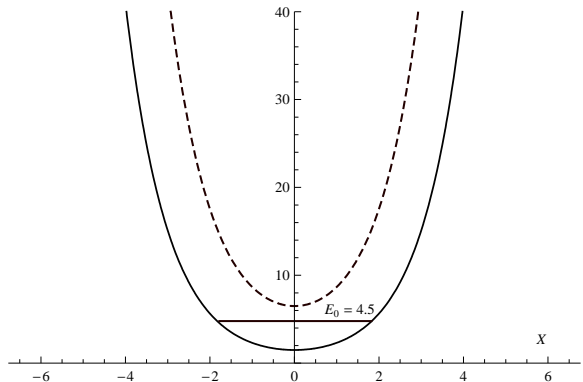}}
\label*{}\caption{Plot of the potential $V(x)$ (solid line) given
in (\ref{e10}) and its first order isospectral partners
$\bar{V}(x)$ (dashed line) given in (\ref{e28}) by deleting the
ground state $E_0=4.5$, we have considered here $a=5, b=0, c=3,
\alpha=1,\beta=0.$ }
\end{figure}

{\bf Strictly isospectral potentials :} The strictly (strict in
the sense that the spectrum of the initial potential and its
isospectral potential are exactly the same) isospectral potentials
can be generated with the help of those seed solutions which
vanish at one of the ends of the $x$-domain.  Now for $\beta=0~
\mbox{and}~ \alpha >0$, it is seen from (\ref{e22}) that
$\mathcal{U}$ is unbounded at $x\rightarrow\infty$ if $|a+b-c|>1$.
But the solution (\ref{e14}) become unbounded for $a+b-c<-1$. So
we must take $a+b-c>1.$ On the other hand
 from (\ref{e23}) it is observed that  $\mathcal{U}(x)\rightarrow 0$ at $x\rightarrow-\infty$ if
$c<2$ or $c>0$ but $\psi_n(x)$ are not normalizable for the values
of $c<2$ so we must take $c>0$. So ~$\mathcal {U}(x)$ vanishes at
$x\rightarrow-\infty$
 and unbounded at $x\rightarrow\infty$~if $a+b-c>1 ~\mbox{and}~ c>0.$ In this case the spectrum of
the isospectral potential as well as original potential are
identical i.e. $E_n=\bar{E}_n~,n=0,1,2...$ Considering the seed
solution as
  $$\mathcal{U}(x)=\frac{e^{\frac{c}{2}x}}{(1+e^{
x})^{\frac{a+b+1}{2}}}~ _2F_1\left(a,b,c,\frac{e^{ x}}{1+e^{
x}}\right),~~~~~a+b-c>1, c>0$$ we have calculated the explicit
form of the partner potential using (\ref{e70}) as
  \begin{equation}\begin{array}{llll}
  \displaystyle \bar{V}(x)=\frac{1}{8}[2c(c-2) e^{-x}+2((a+b-c)^2-1)e^{x}\\~~~~~~~~-
  \frac{sech^2(\frac{x}{2})}{c^2(1+c)~ _2F_1\left(a,b,c,\frac{e^x}{1+e^x}\right)}
  \{-4a^2b^2(1+c) \left(_2F_1\left(1+a,1+b,1+c,\frac{e^x}{1+e^x}\right)\right)^2 + 4abc~ _2F_1\left(a,b,c,\frac{e^x}{1+e^x}\right)\\
  ~~~~\left((a+1)(b+1)
   \left(_2F_1\left(2+a,2+b,2+c,\frac{e^x}{1+e^x}\right)\right)^2
  -(1+c)sinh x \left(_2F_1\left(a+1,b+1,c+1,\frac{e^x}{1+e^x}\right)\right)^2\right)\\
  ~~~~~~~~~~~~~~~~ -8c^2(1+c)cosh^3 x
  \left(_2F_1\left(a,b,c,\frac{e^x}{1+e^x}\right)\right)^2
  \left((a+b)cosh(\frac{x}{2})+(1+a+b-2c) sinh(\frac{x}{2})\right)\}]
  \end{array}
  \end{equation}
 In particular for $a=3, b=5, c=4, \alpha=1, \beta=0$ and using (\ref{e10}), (\ref{e70}) we have obtained
 \begin{equation}\begin{array}{ll}
 \displaystyle V(x)=\frac{1}{4}(23cosh x+7 sinh x)\\
 \bar{V}(x)=\frac{15e^{-x}}{4}+2e^x+\frac{4+e^x-3e^{2x}}{(4+3e^x)^2}
 \end{array}\label{e74}
 \end{equation}
 respectively, which are plotted in figure 2.
 In this case eigenfunctions and eigenvalues of the above partner potential $\bar{V}(x)$ are given by
 \begin{equation}\begin{array}{ll}
\displaystyle \bar{\psi}_n(x)\propto
\frac{e^{3x}\left((2+e^x)(n+7)\mathcal{P}_{n-1}^{(4,4)}\left(-tanh
\left(\frac{x}{2}\right)\right)+(1+e^x)(5+3e^x)\mathcal{P}_n^{(3,3)}(-tanh
(\frac{x}{2}))\right)}{(1+e^x)^6(2+e^x)sech
\left(\frac{x}{2}\right)}\\
\displaystyle \bar{E}_n=n^2+8n+10~,~~~~~n=0,1,2...
\end{array}
 \end{equation}
respectively.
\begin{figure}[h]
\epsfxsize=2.75 in \epsfysize=2 in \centerline{\epsfbox{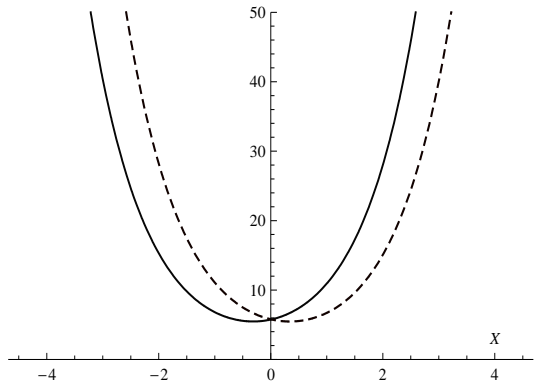}}
\label*{}\caption{Plot of the potential $V(x)$ (solid line) and
its first order isospectral partner (dashed line) given in
(\ref{e74}). }
\end{figure}\\
{\bf Creation of a new ground state :} In this case we shall
consider $\mu<E_0$. The new state can be created below the ground
state of the initial potential with the help of those seed
solutions which satisfies the following two conditions: (i) it
should be nodeless throughout the $x$-domain and (ii) it should be
unbounded at both the end points of the domain of definition of
the given potential $V(x)$. From the asymptotic behaviors of the
seed solution $\mathcal{U}$, given in equations (\ref{e22}) and
(\ref{e23}) we have, for $|a+b-c|>1$ together with either
$c<0~\mbox{or}~c>2$, the above two conditions are satisfied. But
to get $\psi_n(x)$ as physically acceptable, we shall take
$c>2~\mbox{and}~a+b-c>1.$ In this case the spectrum of the partner
potential is $\{\mu, E_n,n=0,1,2...\}$, $E_n$ being the energy
eigenvalues of the original potential $V(x)$ given in (\ref{e10}).
Corresponding bound state wavefunctions are
$\{\bar{\psi}_{\mu}(x), \bar{\psi}_n(x),n=0,1,2,...\}$, where
$\bar{\psi}_{\mu}$ and $\bar{\psi}_n$ are given by (\ref{e72}) and
(\ref{e21}) respectively.\\ For $a+b-c>1, c>2$ and  the seed
solution $\mathcal{U}$ given in (\ref{e13}), the general
expression of the isospectral potential becomes too involved so
instead of giving the explicit expression of the partner potential
we have plotted in figure 3 the original potential $V(x)$ given in
(\ref{e10}) and its partner potential $\bar{V}(x)$ (which is
obtained using (\ref{e70})) considering the particular values $a=
2.8, b=20, c=4.4~ \mbox{and} ~\alpha=\beta=1$. In this case the
energy eigenvalues of $\bar{V}(x)$ are given by
\begin{equation}
\bar{E}_n=\{-13.32,E_n,n=0,1,2...\}=\{-13.32,n^2+22.8n+42.68,n=0,1,2...\}
\end{equation}
Corresponding eigenfunctions can be obtained using the formulae
(\ref{e72}) and (\ref{e21}).
\begin{figure}[h]
\epsfxsize=4 in \epsfysize=2.25 in \centerline{\epsfbox{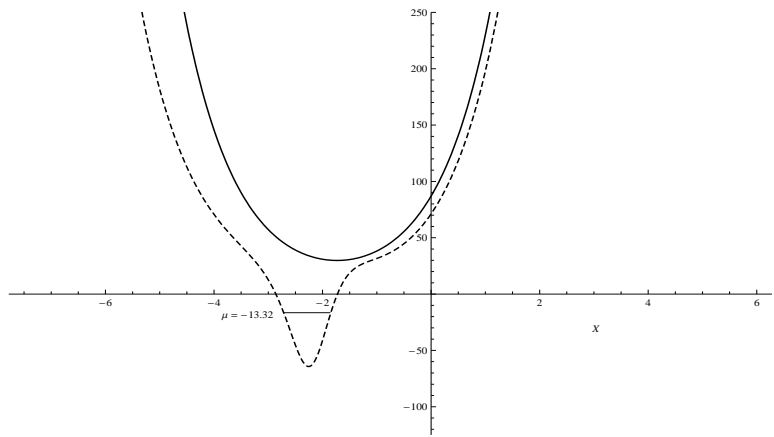}}
\label*{}\caption{Plot of the potential $V(x)$ (solid line) given
in (\ref{e10}) and its first order isospectral partner $\bar{V}$
(dashed line) by inserting the state $\mu=-13.32$. We have
considered here $a=2.8,b=20, c=4.4, \alpha=\beta=1$.}
\end{figure}
\newpage
\section{\label{}  Second order Intertwining} Now we assume the
existence of a second order intertwining operator
\begin{equation}
\mathcal{L}=\frac{1}{M}\frac{d^2}{dx^2}+\eta(x)\frac{d}{dx}+\gamma(x)\label{e37}
\end{equation}
where $\eta(x), \gamma(x)$ are to be determined. Substitution of
this intertwiner in equation (\ref{e60}) and comparison of the
coefficients of like order derivatives leads to a set of following
equations
\begin{equation} \hspace{-7 cm} \bar{V} =
V+2\eta'+\frac{M'}{M}\eta-\frac{3M'^2}{M^3}+\frac{2M''}{M^2}\label{e38}
\end{equation}
\begin{equation}
(\bar{V}-V)\eta =
\frac{2V'}{M}+\frac{2\gamma~'}{M}+\frac{\eta''}{M}-\frac{\eta'M'}{M^2}
+\frac{M''\eta}{M^2}-\frac{2M'^2\eta}{M^3}+\frac{6M'^3}{M^5}-\frac{6M'M''}{M^4}+\frac{M'''}{M^3}\label{e39}
\end{equation}
\begin{equation}
\hspace{-6.8 cm}(\bar{V}-V)\gamma =
\frac{V''}{M}+V'\eta+\frac{\gamma~''}{M}-\frac{M'\gamma~'}{M^2}\label{e40}
\end{equation}
Now using (\ref{e38}) the equations (\ref{e39}) and (\ref{e40})
reads
\begin{equation}\begin{array}{ll} \displaystyle
 2\eta\eta'+\frac{M'\eta^2}{M}-\frac{3\eta
M'^2}{M^3}+\frac{2M''\eta}{M^2}-\frac{2\gamma~'}{M}+\frac{M'\eta'}{M^2}\\
\displaystyle +\frac{2M'^2\eta}{M^3}-\frac{\eta''}{M}
-\frac{M''\eta}{M^2}-\frac{2V'}{M}-\frac{6M'^3}{M^5}+\frac{6M'M''}{M^4}-\frac{M'''}{M^3}=0
 \label{e41}
 \end{array}
\end{equation}
and
\begin{equation}
\gamma\left(2\eta'+\frac{M'\eta}{M}-\frac{3M'^2}{M^3}+\frac{2M''}{M^2}\right)+\frac{M'\gamma~'}{M^2}-\frac{\gamma~''}{M}-\eta
V'-\frac{V''}{M}=0\label{e43}
\end{equation} respectively. Equation (\ref{e41}) can be integrated to obtain
\begin{equation}
\gamma=\frac{M\eta^2}{2}+\frac{M'\eta}{2M}-\frac{\eta'}{2}-V+\frac{M'^2}{M^3}-\frac{M''}
{2M^2}+C_1\label{e42}
\end{equation}
where $C_1$ is an arbitrary constant. Using (\ref{e42}) in
(\ref{e43}) we obtain
\begin{equation}
\begin{array}{lll}
\displaystyle
\frac{\eta'''}{2M}+M\eta'\eta^2-\frac{\eta\eta'M'}{2M}-2\eta'^2-2\eta'V+\frac{5\eta'M'^2}{M^3}-\frac{3\eta'M''}{M^2}
+\frac{\eta^3M'}{2}-\frac{\eta^2M'^2}{2M^2}-\frac{\eta VM'}{M}-\frac{2\eta M'^3}{M^4}\\
\displaystyle +\frac{\eta^2M''}{2M}+\frac{5\eta
M'M''}{2M^3}-\frac{\eta''M'}{M^2}-\eta \eta''-\eta V'-\frac{\eta
M'''}{2M^2}+2C_1\eta'+\frac{C_1\eta
M'}{M}+\frac{3VM'^2}{M^3}-\frac{18M'^4}{M^6}\\
\displaystyle
+\frac{49M'^2M''}{2M^5}-\frac{3C_1M'^2}{M^3}-\frac{2VM''}{M^2}-\frac{4M''^2}{M^4}+\frac{2C_1M''}{M^2}-\frac{V'M'}{M^2}
-\frac{9M'M'''}{2M^4}+\frac{M''''}{2M^3}=0\label{e44}
\end{array}
\end{equation}
Multiplying  by $\left(\eta M+\frac{M'}{M}\right)$, above equation
(\ref{e44}) can be integrated to obtain
\begin{equation}
\begin{array}{lll}
\displaystyle \frac{\eta
\eta''}{2}-\frac{\eta'^2}{4}-\eta'\eta^2M+\frac{\eta^4M^2}{4}-M\eta^2
V+C_1M\eta^2+\frac{C_1M'^2}{M^3}+\frac{2C_1M'\eta}{M}-\frac{2M'V\eta}{M}-\frac{M'^2V}{M^3}-\frac{M''\eta^2}{2M}\\
\displaystyle +\frac{M'\eta^3}{2}
+\frac{5M'^3\eta}{m^4}-\frac{2M'\eta\eta'}{M}+\frac{5M'^2\eta^2}{4M^2}+\frac{M'\eta''}{2M^2}-\frac{M''\eta'}{2M^2}
+\frac{M'''\eta}{2M^2}-\frac{4M'M''\eta}{M^3}+\frac{3M'^4}{M^6}-\frac{5M'^2M''}{2M^5}\\
\displaystyle -\frac{M''^2}{4M^4}+\frac{M'M'''}
{2M^4}+C_2=0\label{e45}
\end{array}
\end{equation}
where $C_2$ is the constant of integration. For a given potential
$V(x)$, the new potential $\bar{V}(x)$ and $\gamma(x)$ can be
obtained from (\ref{e38}) and (\ref{e42}) if the solution
$\eta(x)$ of (\ref{e45}) is known. To obtain $\eta(x)$ we take the
Ans\"{a}tz
\begin{equation}
\eta'=M\eta^2+2\left(\eta+\frac{M'}{M^2}\right)\tau+\frac{M'}{M}\eta+\frac{2M'^2}{M^3}-\frac{M''}{M^2}+\xi\label{e46}
\end{equation}
where $\xi$ is a constant to be determined and $\tau$ is a
function of $x$. Using above ans\"{a}tz in equation (\ref{e45}) we
obtain the following equation
\begin{equation}\begin{array}{ll}\displaystyle
M\left(\frac{\tau'}{M}+\frac{\tau^2}{M}-\frac{M'\tau}{M^2}-V+C_1-\frac{\xi}{2}\right)\eta^2
+\frac{2M'}{M}\left(\frac{\tau'}{M}+\frac{\tau^2}{M}-\frac{M'\tau}{M^2}-V+C_1-\frac{\xi}{2}\right)\eta\\
\displaystyle
+\frac{M'^2}{M^3}\left(\frac{\tau'}{M}+\frac{\tau^2}{M}-\frac{M'\tau}{M^2}-V+C_1-\frac{\xi}{2}\right)
+\left(C_2-\frac{\xi^2}{4}\right)=0\label{e47}
\end{array}
\end{equation}
Since equation (\ref{e47}) is valid for arbitrary $\eta$, the
coefficients of each power of $\eta$ must vanish, which give
$\xi^2=4C_2$ and
\begin{equation}
\frac{\tau'}{M}+\frac{\tau^2}{M}-\frac{M'\tau}{M^2}-V+C_1-\frac{\xi}{2}=0\label{e48}
\end{equation}
Now defining $\mu=C_1-\frac{\xi}{2}$ , the above equation can be
written as
\begin{equation}
\frac{\tau'}{M}+\frac{\tau^2}{M}-\frac{M'\tau}{M^2}=V-\mu~,~~~~\mu=C_1-\frac{\xi}{2}=C_1\mp\sqrt{C_2}\label{e49}
\end{equation}
The equation (\ref{e49}) is a Riccati equation which can be
linearized by defining $\tau=\frac{\mathcal{U'}}{\mathcal{U}}$.
Making this change in equation (\ref{e49}) we obtain
\begin{equation}
-\frac{1}{M}\mathcal{U}''-\left(\frac{1}{M}\right)'\mathcal{U}'+V\mathcal{U}=\mu
\mathcal{U}\label{e54}
\end{equation}
Depending on whether $C_2$ is zero or not, $\xi$ vanishes or takes
two different values $\pm \sqrt{C_2}$. If $C_2 = 0$, we need to
solve one equation of the form (\ref{e49}) and then the equation
(\ref{e46}) for $\eta (x)$. If $C_2 \neq 0$, there will be two
different equations of type (\ref{e49}) for two factorization
energies $\mu_{1,2}=C_1\mp\sqrt{C_2}$. Once we solve them, it is
possible to construct algebraically a common solution $\eta (x)$
of the corresponding pair of equations (\ref{e46}). There is an
obvious difference between the real case with $C_2 > 0$ and the
complex case $C_2 < 0$; thus there follows a natural scheme of
classification for the solutions $\eta (x)$ based on the sign of $C_2$.
 In our present article we shall not discuss the case $C_2=0.$ \\

{\bf (i) Real Case $(C_2>0)$}\\
Here we have $\mu_{1,2}\in\mathbb{R}, \mu_1\neq\mu_2$. Let the
corresponding solutions of the  Riccati  equation (\ref{e49}) be
denoted by $\tau_{1,2}(x).$ Now the associated pair of equations
(\ref{e46}) become
\begin{equation}
\eta'=M\eta^2+2\left(\eta+\frac{M'}{M^2}\right)\tau_1+\frac{M'}{M}\eta+\frac{2M'^2}{M^3}-\frac{M''}{M^2}+\mu_2-\mu_1\label{e76}
\end{equation}
and
\begin{equation}
\eta'=M\eta^2+2\left(\eta+\frac{M'}{M^2}\right)\tau_2+\frac{M'}{M}\eta+\frac{2M'^2}{M^3}-\frac{M''}{M^2}+\mu_1-\mu_2\label{e77}
\end{equation}
respectively. Subtracting (\ref{e76}) from (\ref{e77}) and using
(\ref{e54}) we obtain $\eta(x)$ as
\begin{equation}
\eta(x)=\frac{\mu_1-\mu_2}{\tau_1-\tau_2}-\frac{M'}{M^2}=-\frac{W'(\mathcal{U}_1,\mathcal{U}_2)}{M
W(\mathcal{U}_1,\mathcal{U}_2)}\label{e55}
\end{equation}
where $\mathcal{U}_1 , \mathcal{U}_2$ are the seed solutions of
the equation (\ref{e54}) corresponding to the factorization energy
$\mu_1$ and $\mu_2$ respectively and
$W(\mathcal{U}_1,\mathcal{U}_2)=\mathcal{U}_1\mathcal{U}'_2-\mathcal{U}'_1\mathcal{U}_2$,
is the Wronskian of
$\mathcal{U}_1$ and $\mathcal{U}_2$. \\
Now it is clear from (\ref{e38}) and (\ref{e55}) that  mass
function $M(x)$ is nonsingular and does not vanish at the finite
part of the $x$-domain, so that the new potential $\bar{V}(x)$ has
no extra singularities (i.e. the number of singularities in $V$
and $\bar {V}$ remains the same) if
$W(\mathcal{U}_1,\mathcal{U}_2)$ is nodeless there. The spectrum
of $\bar{\mathcal{H}}$ depends on whether or not its two
eigenfunctions $\bar{\psi}_{\mu_{1,2}}$ which belongs as well to
the kernel of $\mathcal{L}^{\dagger}$ can be normalized
\cite{r13}, namely
\begin{equation}\mathcal{L}^{\dagger}\bar{\psi}_{\mu_{j}}=0~~~~\mbox{and}~~~~
\bar{\mathcal{H}}\bar{\psi}_{\mu_{j}}=\mu_{j}\bar{\psi}_{\mu_{j}},~~~j=1,2\label{e56}
\end{equation}
where $\mathcal{L}^{\dagger}$ is the adjoint of $\mathcal{L}$ and
is given by \cite{r22}
$$\mathcal{L}^{\dagger}=\frac{1}{M}\frac{d^2}{dx^2}-\left(\eta+\frac{2M'}{M^2}\right)\frac{d}{dx}+\left(\frac{2M'^2}{M^3}-\frac{M''}{M^2}
-\eta'+\gamma\right)$$ For $j=1$ the explicit expression of the
two equation mentioned in (\ref{e56}) are
\begin{equation}
\frac{1}{M}\frac{d^2\psi_{\mu_1}}{dx^2}-\left(\eta+\frac{2M'}{M^2}\right)\frac{d\psi_{\mu_1}}{dx}+\left(\frac{2M'^2}{M^3}-\frac{M''}{M^2}
-\eta'+\gamma\right)\psi_{\mu_1}=0\label{e78}
\end{equation}
and
\begin{equation}
-\frac{1}{M}\psi_{\mu_1}''-\left(\frac{1}{M}\right)'\psi_{\mu_1}'+(\bar{V}-\mu_1)\psi_{\mu_1}=0\label{e79}
\end{equation}
respectively. Adding (\ref{e78}) from (\ref{e79}) we obtain
\begin{equation}
-\left(\frac{M'}{M^2}+\eta\right)
\frac{d\psi_{\mu_1}}{dx}+\left(\bar{V}-\mu_1+\frac{2M'^2}{M^3}-\eta'+\gamma-\frac{M''}{M^2}\right)\psi_{\mu_1}=0\label{e80}
\end{equation}
Substituting the values of $\bar{V}$ and $\gamma$ from (\ref{e38})
and (\ref{e42}) with $2C_1=\mu_1+\mu_2$, in the above equation
(\ref{e80}), we get
\begin{equation}
\frac{d}{dx}\left(log{\psi_{\mu_1}}\right)=\frac{\eta'+3\eta\frac{M'}{M}+\frac{M''}{M^2}+M\eta^2+2(C_1-\mu_1)}{2(\eta+\frac{M'}{M^2})}\label{e81}
\end{equation}
Now using our ans\"{a}tz (\ref{e46}) in (\ref{e81}) and then
integrating we obtain
\begin{equation}
\psi_{\mu_1}\propto\frac{M\left(\eta+\frac{M'}{M^2}\right)}{\mathcal{U}_1}\propto\frac{M\mathcal{U}_2}{W(\mathcal{U}_1,\mathcal{U}_2)}\label{e82}
\end{equation}
Above procedure can be applied to obtain  $\bar{\psi}_{\mu_{2}}$
as
\begin{equation}
\bar{\psi}_{\mu_2}\propto\frac{\eta
M+\frac{M'}{M}}{\mathcal{U}_2}\propto\frac{M\mathcal{U}_1}{W(\mathcal{U}_1,\mathcal{U}_2)}\label{e83}
\end{equation}
If both $\bar{\psi}_{\mu_{1,2}}$ are normalizable then we get the
maximal set of eigenfunctions of $\bar{\mathcal{H}}$
as$\{\bar{\psi}_{\mu_1},\bar{\psi}_{\mu_2},\bar{\psi}_{n}\propto\mathcal{L}\psi_n\}$.
Among the several spectral modifications which can be achieved
through the real second order SUSYQM for PDMSE,
 some cases are worth to be mentioned.\\

 {\bf{ Deletion of first two energy levels :}} For $\mu_1=E_0$
and $\mu_2=E_1$ the two solutions of equation (\ref{e54}) are the
normalizable solutions of equation (\ref{e15}) i.e,
$\mathcal{U}_1=\psi_0(x)$ and $\mathcal{U}_2=\psi_1(x)$
respectively. It turns out that the Wronskian is nodeless but two
solutions $\bar{\psi}_{\mu_1}$ and $\bar{\psi}_{\mu_2}$ are
non-normalizable. Thus
$Sp(\bar{\mathcal{H})}=Sp(\mathcal{H})-\{E_0,E_1\}=\{E_2,E_3,E_4,...\}$,
i.e., the two levels
$E_0$ and $E_1$ are deleted to generate $\bar{V}.$\\

 {\bf Isospectral transformations :}
If we take $\mu_1 < \mu_2<E_0$ and choose  $\mathcal{U}_1$ and
$\mathcal{U}_2$ such way that either $\mathcal{U}_{1,2}(x_l)=0$ or
$\mathcal{U}_{1,2}(x_r)=0$, $x_l$ and $x_r$ being the end points
of the domain of definition of $V(x)$, then the Wronskian
$W(\mathcal {U}_1,\mathcal {U}_2)$ vanishes at $x_l$ or $x_r$.
Hence $\bar\psi_{\mu_1}$ and $\bar{\psi}_{\mu_2}$ become
non-normalizable so that
$Sp(\mathcal{H})=Sp(\mathcal{\bar{H}})$.\\

{\bf{ Creation of two new levels below the ground state :}} For
$\mu_2 < \mu_1<E_0$ and choosing $\mathcal{U}_1$ and
$\mathcal{U}_2$ in such way that $\mathcal{U}_2$ has exactly one
node and $\mathcal{U}_1$ is nodeless then the Wronskian
$W(\mathcal {U}_1,\mathcal {U}_2)$ becomes nodeless, also two
wavefunctions $\bar{\psi}_{\mu_1}$ and $\bar{\psi}_{\mu_2}$ are
normalizable. Therefore the spectrum of $\mathcal{\bar{H}}$
becomes
$Sp(\bar{\mathcal{H}})=Sp(\mathcal{H})\bigcup\{\mu_1,\mu_2\}=\{\mu_1,\mu_2,E_n,n=0,1,2...\}$
i.e. two new levels have been inserted
to the spectrum of $V(x)$ to obtain $\bar{V}(x).$\\

{\bf{(ii) Complex case $(C_2<0)$}}\\
For $C_2<0$ the two factorization energies $\mu_1$ and $\mu_2$
become complex. In order to construct real $\bar{V}$ we shall
choose $\mu_1$ and $\mu_2$ as complex conjugate to each other i.e,
$\mu_1=\mu\in \mathbb{C}$ and $\mu_2=\bar{\mu}$. For the same
reason we shall take $\tau_1(x)=\tau(x)$ and
$\tau_2(x)=\bar{\tau}(x).$ Hence the real solution $\eta(x)$ of
(\ref{e46}) generated from the complex $\tau(x)$ of (\ref{e49})
becomes
\begin{equation}
\eta(x)=\frac{\mu-\bar{\mu}}{\tau-\bar{\tau}}-\frac{M'}{M^2}=\frac{Im(\mu)}{Im(\tau)}-\frac{M'}{M^2}
=-\frac{W'(\mathcal{U},\bar{\mathcal{U}})}
{MW(\mathcal{U},\bar{\mathcal{U}})}\label{e91}
\end{equation}
Defining $w(x)=\frac{W(\mathcal {U},\bar{\mathcal
{U}})}{M(\mu-\bar{\mu})}$, $\eta(x)$ becomes
\begin{equation}
\eta(x)=-\frac{w'}{Mw}-\frac{M'}{M^2}
\end{equation}
For the factorization energies $\mu$ and $\bar{\mu}$ the equation
(\ref{e54}) becomes
$$-\frac{1}{M}\mathcal{U}''-\left(\frac{1}{M}\right)'\mathcal{U}'+V\mathcal{U}=\mu~
\mathcal{U} ~~\mbox{and}~~
-\frac{1}{M}\mathcal{\bar{U}}''-\left(\frac{1}{M}\right)'\mathcal{\bar{U}}'+V\mathcal{\bar{U}}=\bar{\mu}~
\mathcal{\bar{U}}$$ Multiplying first equation by
$\mathcal{\bar{U}}$ and second equation by $\mathcal{U}$ and then
subtracting we obtain
\begin{equation}
\frac{W'(\mathcal{U},\mathcal{\bar{U}})}{M(\mu-\bar{\mu})}-\frac{M'W(\mathcal{U},\mathcal{\bar{U}})}{M^2(\mu-\bar{\mu})}=|\mathcal{U}|^2\label{e84}
\end{equation}
Using above relation (\ref{e84}) we have
\begin{equation}
w'(x)=\frac{W(\mathcal{U},\mathcal{\bar{U}})}{M(\mu-\bar{\mu})}-\frac{M'W(\mathcal{U},\mathcal{\bar{U}})}{M^2(\mu-\bar{\mu})}=|\mathcal{U}|^2
\end{equation}
which implies that $w(x)$ is a non-decreasing function. So it is
sufficient to choose
\begin{equation}
\lim_{x\rightarrow x_l} \mathcal{U}=0 ~\mbox{or}
~~\lim_{x\rightarrow x_r} \mathcal{U}=0
\end{equation}
for the Wronskian $W$ to be nodeless. It is to be noted here that
in this case we can only construct potentials which are strictly
isospectral with the initial potential.\\
\subsection{Second-order intertwining and type A 2-fold SUSY}
The second-order intertwiner ${\cal{L}}$ in equation (\ref{e37})
coincides with $P_2$ given in equation (\ref{e103}) if one takes
\begin{equation}
\eta(x)= \frac{2 W(x)}{M(x)}- \frac{M'(x)}{M(x)^2}\label{e108}
\end{equation}
It is now easy to verify that for this $\eta(x)$, $(V^+-V^-)$
given in (\ref{e104}) (with ${\cal{N}}=2$) agree with
$(\bar{V}-V)$ given in (\ref{e38}).

Now it is to be shown that the Hamiltonian $\cal{H}$ given in
equation (\ref{e15}) admits two eigenfunctions
~${\cal{U}}_{1,2}(x)$ corresponding to two factorization energies
$\mu_{1,2}$ respectively i.e.,
\begin{equation}
{\cal{H~U}}_i(x) = \mu_i ~{\cal{U}}_i(x),~~~i=1,2\label{e109}
\end{equation}
will belong to type A 2-fold SUSY in PDM background (in constant
mass scenario this was already proved in ref.\cite{r28}). For this
we define
\begin{equation}
z(x)= \frac{{\cal{U}}_2(x)}{{\cal{U}}_1(x)},~~~~~~
{\cal{W}}_{2}^-(z)\equiv{\cal{W}}(z)=-ln
~{\cal{U}}_1(x)\label{e110}
\end{equation}
For this ${\cal{W}}(z)$, it is evident that the gauged Hamiltonian
$\tilde{\cal{H}}_2^-$ defined by
\begin{equation}
\tilde{\cal{H}}_2^- = e^{\cal{W}}{\cal{H}}e^{-\cal{W}}\label{e111}
\end{equation}
must be diagonal in the basis $\tilde{\nu}_2= \left\langle
1,z\right\rangle$ because of the assumption (\ref{e109}) and the
choice (\ref{e110}). From equation (\ref{e105}), its immediate
consequence (for ${\cal{N}}=2$) is
\begin{equation}
B(z)=\frac{z''(x)}{M(x)}
-\frac{z'(x)M'(x)}{M(x)^2}-\frac{2z'(x)^2}{M(x)}\frac{d{\cal{W}}(z)}{dz}\label{e112}
\end{equation}
Using equations (\ref{e110}) and (\ref{e109}) in the above
equation (\ref{e112}) it can be shown that
\begin{equation}
B(z)=(\mu_1-\mu_2)~ z(x)\label{e113}
\end{equation}
For this value of $B(z)$ it is also easy to verify that the
expression of $\eta(x)$ in equation (\ref{e108}) and (\ref{e55})
are same. Now it is evident that the gauged Hamiltonian
${\tilde{H}}_2^-$ preserves the vector space
$\tilde{\nu}_2=\left\langle 1,z\right\rangle$. Hence it is
possible to get type A 2-fold SUSY system
$(\cal{H},\bar{\cal{H}},\cal{L})$ following the prescription given
in ref. \cite{r27}, with the choice of $z(x)$, ${\cal{W}}(z)$ and
$\tilde{\cal{H}}_2^-$ given by (\ref{e110}) and (\ref{e100})
respectively.
\subsection{Example of second-order intertwining for real factorization energies}
It may be emphasized that the results mentioned in section III are
most general and valid for any potential $V(x)$. However to
illustrate the above procedure with the help of an example we
shall need non-normalizable solutions of (\ref{e54}) (which is
similar to equation (\ref{e18}) but with two factorization
energies) corresponding to a particular mass function $M(x)$. To
illustrate the second order intertwining with an example we have
considered the potential (\ref{e10}) as an initial potential.
 Corresponding seed solution for the factorization energy $\mu =
 \mu_1$ which is obtained in Appendix A, is
\begin{equation}
\mu_1=-ab+\frac{(a+b+1)c}{2}-\frac{c^2}{2}\label{e75}
\end{equation}
\begin{equation}
 \mathcal{U}_1(x)=\alpha\frac{e^{\frac{c}{2}x}}{(1+e^{
x})^{\frac{a+b+1}{2}}}~ _2F_1\left(a,b,c,\frac{e^{ x}}{1+e^{
x}}\right) +\beta\frac{e^{\left(1-\frac{c}{2}\right)
x}}{(1+e^{x})^{\frac{a+b-2c+3}{2}}}~
 _2F_1\left(a-c+1,b-c+1,2-c,\frac{e^{ x}}{1+e^{ x}}\right)\label{e50}
 \end{equation}
We notice that the potential (\ref{e10}) and corresponding
Hamiltonian are invariant under the transformation $a\rightarrow
a+\nu$ and $b\rightarrow b-\nu$, $\nu\in\mathbb{R}-\{0\}$. But the
solution (\ref{e13}) of the corresponding Schr\"{o}dinger equation
changes to
\begin{equation}
 \begin{array}{ll}
 \displaystyle
 \mathcal{U}_2(x)=\alpha\frac{e^{\frac{c}{2}x}}{(1+e^{
x})^{\frac{a+b+1}{2}}}~ _2F_1\left(a+\nu,b-\nu,c,\frac{e^{
x}}{1+e^{
x}}\right)\\
\displaystyle ~~~~~~~~~~~~~~~~~~~~~~~~~~
+\beta\frac{e^{\left(1-\frac{c}{2}\right) x}}{(1+e^{
x})^{\frac{a+b-2c+3}{2}}}~
 _2F_1\left(a+\nu-c+1,b-\nu-c+1,2-c,\frac{e^{ x}}{1+e^{
 x}}\right)\label{e51}
 \end{array}
 \end{equation}
 and the corresponding factorization energy is given by
 \begin{equation}
\mu_2=-ab+\frac{(a+b+1)c}{2}-\frac{c^2}{2}+\nu(a-b)+\nu^2\label{e52}
 \end{equation}
 Thus the general solutions of the equation (\ref{e54}) for the two
 factorization energies $\mu_1$ and $\mu_2$, are given by (\ref{e50}) and
 (\ref{e51})respectively.
 The asymptotic behaviors  of the seed solution $\mathcal{U}_2$ remains same as $\mathcal{U}_1$, which are given in
  (\ref{e22}) and (\ref{e23}).\\
{\bf Deletion of  first two energy levels :}
 Let us take $\mu_1=E_0$ and $\mu_2=E_1$,
 $\mathcal{U}_1=\psi_0(x)$
 and $\mathcal{U}_2=\psi_1(x)$ which are given in (\ref{e14}). The Wronskian
 $W(\mathcal{U}_1,\mathcal{U}_2)$ is given by
 \begin{equation}
 W(\mathcal{U}_1,\mathcal{U}_2)\propto \frac{e^{(c+1)x}}{(1+e^x)^{a+b+3}}\label{e53}
 \end{equation}
which is nodeless and bounded in $(-\infty,\infty)$ as
$c>-\frac{1}{2}$ and $a+b-c+\frac{1}{2}>0$ (these conditions are
mentioned at the end of the Appendix A). The second-order SUSY
partner of $V(x)$ is obtained using equation (\ref{e38}) and is
given by
\begin{equation}
\bar{V}(x)=\frac{1}{4}\left[\left(c^2-2c(a+b+2)+(a+b+1)(a+b+3)\right)e^x+c(c+2)e^{-x}+4(a+b+1)\right]\label{e85}
\end{equation}
Clearly the eigenfunctions
$\bar{\psi}_{\mu_1}\propto\frac{M\mathcal{U}_2}{W(\mathcal{U}_1,\mathcal{U}_2)}$
and
$\bar{\psi}_{\mu_2}\propto\frac{M\mathcal{U}_1}{W(\mathcal{U}_1,\mathcal{U}_2)}$
of $\bar{\mathcal{H}}$ associated to $\mu_1=E_0$ and $\mu_2=E_1$
are not normalizable since
$$\lim_{x\rightarrow-\infty,\infty}\bar{\psi}_{\mu_{1,2}}(x)=\infty$$
Thus
$Sp(\bar{\mathcal{H}})=Sp(\mathcal{H})-\{E_0,E_1\}=\{E_2,E_3...\}.$\\
In particular taking $a=5,b=0,c=3$ we have plotted the potential
$V(x)$ and its second-order SUSY partner $\bar{V}(x)$ given in
(\ref{e10}) and (\ref{e85}) respectively, in figure 4.
\begin{figure}[h]
\epsfxsize=2.75 in \epsfysize=2 in \centerline{\epsfbox{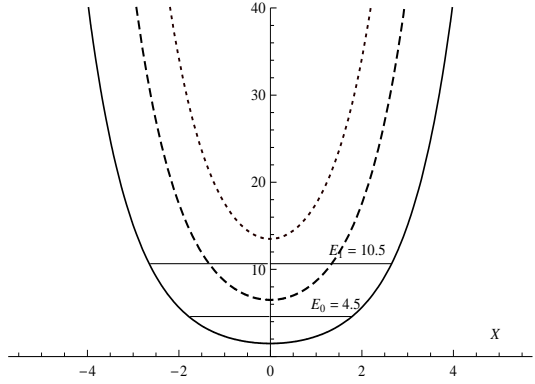}}
\label*{}\caption{ Plot of the original potential (solid line) for
$a= 5, b=0, c=3$ and its first-order SUSY partner (dashed line) by
deleting the ground state $E_0=4.5$ and second-order SUSY partner
(dotted line) by deleting two successive states $E_0=4.5,
E_1=10.5.$}
\end{figure}\\
  {\bf{Strictly isospectral potentials :}}
 The strictly isospectral partner potentials can be constructed by creating two new
 energy levels in the limit when each seed vanishes at one of the
 ends of the $x$-domain. Now from the asymptotic behaviors of
 the seed solutions, we note that both the seed solutions vanish at
 $x\rightarrow -\infty$ for $\beta=0, \alpha>0$ if $a+b-c>1$ and
 $c>0$. Considering $\beta=0, \alpha=1$ in (\ref{e50}) and (\ref{e51}) we take two seed solution as
 \begin{equation}
 \mathcal{U}_1(x)=\frac{e^{\frac{c}{2}x}}{(1+e^{
x})^{\frac{a+b+1}{2}}}~ _2F_1\left(a,b,c,\frac{e^{ x}}{1+e^{
x}}\right)
 \end{equation}
 and
 \begin{equation}
 \mathcal{U}_2(x)=\frac{e^{\frac{c}{2}x}}{(1+e^{
x})^{\frac{a+b+1}{2}}}~ _2F_1\left(a+\nu,b-\nu,c,\frac{e^{
x}}{1+e^{ x}}\right)
 \end{equation}
 Since $\mathcal{U}_{1,2}(x)\rightarrow 0$ at $x\rightarrow -\infty$, from the expressions (\ref{e82}) and
 (\ref{e83}) we can conclude that
  $$\lim_{x\rightarrow -\infty}\bar{\psi}_{\mu_{1,2}}(x)=\infty$$ which
 implies that $\mu_{1,2}$ does not belongs to
 $Sp(\bar{\mathcal{H}})$ i.e. $\bar{V}(x)$ is strictly isospectral
 to $V(x).$ Here the general expression of the partner potential is too involved so instead of giving
 the explicit expression we have considered particular values $a=3 , b = 5, c=4, \nu
 =1, \alpha=1, \beta=0$. Corresponding expression of the partner potential and its energy spectrum
 are $$\bar{V}(x)= 1+\frac{3}{4} (9 cosh x- 7 sinh x), ~~~~~E_n=\bar{E}_n=n^2+8 n+10,~~n=0,1,2...$$
 respectively. In figure 5, we have plotted the initial potential,
 its first and second-order strictly isospectral partner
 potentials for the parameter values $a=3,b=5 ,c=4,\alpha=1, \beta=0,\nu=1$.
 \begin{figure}[h]
\epsfxsize=2.75 in \epsfysize=2 in \centerline{\epsfbox{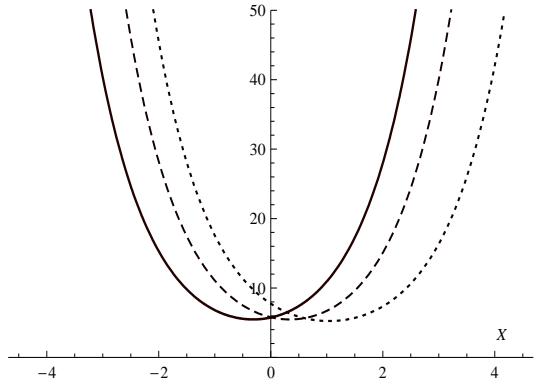}}
\label*{}\caption{ Plot of the original potential (solid line) and
its first-order (dashed line) and second-order (dotted line) SUSY
partner by making the isospectral transformation for $a=3,b=5
,c=4,\alpha=1, \beta=0,\nu=1$.}
\end{figure}
\newpage
{\bf Creation of two new levels below the ground state : } Two
energy levels can be created taking $\mu_2<\mu_1<E_0$ and using
those seed solutions $\mathcal{U}_1$ and $\mathcal{U}_2$ for which
the Wronskian become nodeless. In this case the expressions of the
Wronskian contains several Hypergeometric function, so it is very
difficult to mention the range of $a,b,c$ and $\nu$ for which it
is nodeless. In particular for $a= 2.8, b=20, c=4.4 , \alpha=1,
\beta=1$ we have the Wronskian is found to be nodeless. For the
same values of $a,b,c$ we have plotted the potential and its
second order partner in figure 6. The second-order isospectral
partner is obtained using equation (\ref{e38}).
\begin{figure}[h]
\epsfxsize=4.5 in \epsfysize=2.7 in
\centerline{\epsfbox{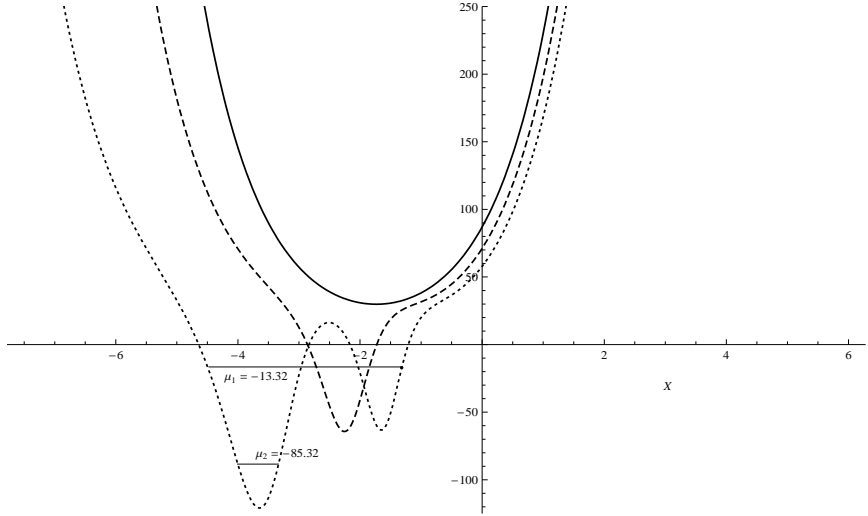}} \label*{}\caption{Plot of the
original potential (solid line) for $a= 2.8, b=20, c=4.4$ and its
first-order SUSY partner (dashed line) by creating a new level
$\mu_1=-13.32$ and second-order SUSY partner (dotted line) by
creating two new levels $\mu_1=-13.32, \mu_2=-85.32$}
\end{figure}
\newpage
\subsection{Example of second order intertwining for complex factorization energies}
As mentioned earlier, in this case we can only construct the
strictly isospectral partner potentials. The complex factorization
energy $\mu_1$ and $\mu_2$ given by equation (\ref{e75}) and
(\ref{e52}), can be made conjugate to each other in several ways.
One of the way is by making following restrictions on $a,b, c,\nu$
{\bf :} $c\in \mathbb{R}, ~Im(a)=-Im(b),~ \nu=Re(b)-Re(a)$. But in
order to keep the initial potential real we have to made two more
restrictions e.g. $Re(a)+Re(b)-c>1$ and $c>2.$ In particular
taking $a= 6.1-5 ~i,b=8+5 ~i,c=4.1 , \nu=1.9$ we have two
factorization energy $ \mu_1(=\mu) =-51.25+9.5~i$ and $
\mu_2(=\bar{\mu}) = -51.25-9.5~i.$  For these values of $a,b,c,
\nu$ and $\alpha=1, \beta=0$ the seed solution $\mathcal{U}$
becomes
\begin{equation}
\mathcal{U}(x)= \frac{e^{2.05 x}}{(1+e^x)^{7.55}}~
_2F_1\left(6.1-5~i,8+5 ~i,4.1,\frac{e^x}{1+e^x}\right)
\end{equation}
Clearly $\mathcal{U}(-\infty)=0$ and $|\mathcal{U}|\rightarrow
\infty$ as $x\rightarrow \infty$ so this seed $\mathcal{U}$ and
its conjugate $\mathcal{\bar{U}}$ can be used to obtain the
second-order SUSY partner potential $\bar{V}(x)$ with the help of
equations (\ref{e38}) and (\ref{e91}). In figure 7 we have plotted
the initial potential $V(x)$  given in (\ref{e10}) and its
isospectral partner $\bar{V}(x)$ for the parameter values
mentioned earlier.
\begin{figure}[h]
\epsfxsize=2.75 in \epsfysize=2 in \centerline{\epsfbox{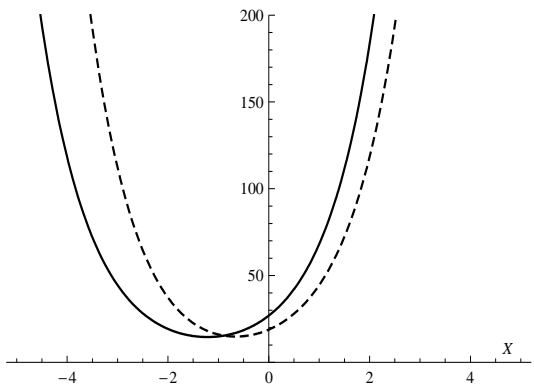}}
\label*{}\caption{ Plot of the original potential (solid line) for
$a= 6.1-5 ~i,b=8+5 ~i,c=4.1, \nu=1.9, \mu_1=-51.25+9.5~i,
\mu_2=-51.25-9.5 ~i.$ and its second-order isospectral partner
(dashed line).}
\end{figure}

\section{ Summary and Outlook}

In this article we have discussed the possibilities for designing
quantum spectra of position dependent mass Hamiltonians offered by
the intertwining technique. For doing this, we start with the
non-normalizable solution of position dependent mass Schr\"odinger
equation with the initial potential (obtained by using the point
canonical transformation approach). To generate spectral
modifications by first order intertwining, we have used solutions
to the position dependent mass Schr\"odinger equation
corresponding to factorization energy (not belonging to the
physical spectrum of the initial problem) less than or equal the
ground state energy in order to avoid singularity in the
isospectral partner potential provided the mass function is not
singular and is not equal to zero in the real line. Thus it is
possible to generate isospectral partner potentials (a) with the
ground state of the original potential deleted (b) with a new
state created below the ground state of the original potential (c)
with the spectrum of the original potential unaffected. In ref
\cite{r10}, the first order intertwining technique was illustrated
by considering the free particle case.

In the case of second order intertwining, instead of using the
iterative method used in \cite{r10}, the second order intertwiner
is constructed directly by taking it as second order linear
differential operator with unknown coefficients which are
functions of $x$. The main advantage of this construction is that
one can generate second-order isospectral partner potentials
directly from the initial potential without generating first-order
partner potentials. The apparently intricate system of equations
arising from the intertwining relationship is solved for the
coefficients by taking an ansatz. In this case the spectral
modifications are done by taking appropriately chosen
factorization energies which may be real or complex. For real
unequal factorization energies, it is possible to generate
potentials (a) with deletion of first two energy levels (b) with
two new levels embedded below the ground state of the original
potential (c) with identical spectrum as of the original
potential. For complex factorization energies, it is shown how to
obtain strictly isospectral potentials. It must be mentioned here
that in all the above cases the conditions for having spectral
modifications remain the same as in the case of constant mass
scenario \cite{r13} provided the mass function $M$ is nonsingular
and is not equal to zero in the finite part of the real line.

In this article, the equivalence of our formalism to type A
${\cal{N}}$-fold $({\cal{N}}=1,2)$ SUSY in PDM background is
shown. Also, it is shown that an arbitrary one body quantum PDM
Hamiltonian which admits two eigenfunctions in closed form belongs
to type A 2-fold SUSY as was previously done in constant mass
scenario \cite{r28}.

Some of the interesting issues to be investigated in future are\\
(i) to obtain spectral changes that appear above the ground state energy of the initial potential.
 Specifically, how to create/delete a pair of levels between any two neighboring initial ones, how to move an arbitrary level or delete an arbitrary level. Specially interesting will be the possibility of embedding a single level at any arbitrary position.\\
(ii) to obtain spectral modifications when the two factorization
energies are equal.

\appendix
\section{Construction of exactly solvable effective potential via PCT}
In order to find the general (unbounded) solution of the equation
(\ref{e18}) we shall use PCT method in PDM background \cite{r14}
to solve this equations. Let us find the solution of equation
(\ref{e18}) of the form
\begin{equation}
\mathcal{U}(x)=f(x)~F(a,b,c,g) \label{e1}
\end{equation}
where $f(x), g(x)$ are two function of $x$ to be determined and
 $F(a,b,c,g)$ is the Hypergeometric  function which satisfies second order differential equation of
 the type
 \begin{equation}
 \frac{d^2F}{dg^2}+Q(g) \frac{dF}{dg}+R(g) F=0,~~\mbox{with}~~~ Q(g)=\frac{c-(a+b+1)g}{g(1-g)}~~~\mbox{and}~~~
 R(g)=-\frac{ab}{g(1-g)}\label{e2}
 \end{equation}
 Substituting equation (\ref{e1}) in (\ref{e18}) we obtain
 \begin{equation}
 \frac{d^2F}{dg^2}+\left(\frac{g''}{g'^2}+\frac{2f'}{fg'}-\frac{M'}{Mg'}\right)
 \frac{dF}{dg}+\left(\frac{f''}{fg'^2}+(\mu-V)\frac{M}{g'^2}-\frac{M'f'}{Mfg'^2}\right)F=0\label{e3}
 \end{equation}
 Comparing equation (\ref{e2}) and (\ref{e3}) we get
 \begin{equation}
\begin{array}{lcl}
Q(g(x)) &=& \frac{g''}{g'^2}+\frac{2f'}{fg'}-\frac{M'}{Mg'}\\
R(g(x)) &=&
\frac{f''}{fg'^2}+(\mu-V)\frac{M}{g'^2}-\frac{M'f'}{Mfg'^2}\label{e4}
\end{array}
\end{equation}
After simplification of the above equation (\ref{e4}) we obtain
\begin{equation}
f(x)\propto \sqrt{\frac{M}{g'}}
~~~~exp\left(\frac{1}{2}\int^{g(x)}Q(t)dt\right)\label{e5}
\end{equation}
and
\begin{equation}
\mu-V=\frac{g'''}{2Mg'}-\frac{3}{4M}\left(\frac{g''}{g'}\right)^2+\frac{g'^2}{M}
\left(R-\frac{1}{2}\frac{dQ}{dg}-\frac{Q^2}{4}\right)-\frac{M''}{2M^2}+\frac{3M'^2}{4M^3}
\label{e6}
\end{equation}
respectively.
 Now in PCT approach there are many options for choosing $M(x)$ \cite{r14},
for example $M(x) = \lambda g'^2(x)$ , $M = \lambda g'(x)$, $M =
\displaystyle \frac{\lambda}{g'(x)}$ , $\lambda$
 being a constant. Here we choose $M(x)=\lambda g'(x)$. For this choice of the mass
 function and using the values of $Q(g), R(g)$ given in equation
 (\ref{e2}),  equation (\ref{e6}) reduces to
 \begin{equation}
 \mu-V=\frac{g'}{\lambda} \left[-\frac{ab}{g(1-g)}-\frac{\left(c-(a+b+1)g\right)^2}{4g^2(1-g)^2}+
 \frac{a+b+1}{2g(1-g)}+\frac{c-(a+b+1)g}{2g^2(1-g)}-\frac{c-(a+b+1)g}{g(1-g)^2}\right]\label{e7}
 \end{equation}
 Now in order to generate a constant term on the right hand side of the above equation which will
  correspond to $\mu$ on the left-hand side, we set
 $\frac{g'}{\lambda g(1-g)}=p $, where $p$ is a positive constant. This
 gives
 \begin{equation}
 g(x)=\frac{e^{p\lambda x}}{1+e^{p\lambda
 x}}~~~\mbox{and}~~~M(x)=\frac{p\lambda^2}{4}sech^2\left(\frac{p\lambda
 }{2}x\right),~~~~~~-\infty< x < \infty\label{e11}
 \end{equation}
 For these values of $g(x)~\mbox{and} ~ M(x)$ we obtain from equation
 (\ref{e7}) new potential $V(x)$ and factorization energy $\mu$ as
 \begin{equation}
V(x)=\frac{[(a+b-c)^2-1]p}{4} e^{p\lambda x}+\frac{cp(c-2)}{4}
e^{-p\lambda x},~~~~~~~-\infty< x < \infty\label{e88}
\end{equation}
and
\begin{equation}
 \mu=-abp+\frac{(a+b+1)cp}{2}-\frac{c^2p}{2}\label{e87}
\end{equation}
Also from (\ref{e5}) we get
$$f(x)=\frac{e^{\frac{cp\lambda}{2}x}}{(1+e^{p\lambda x})^{\frac{a+b+1}{2}}}$$
Hence the solution of the equation (\ref{e18}) at the
factorization energy $\mu$ is given by
\begin{equation}
\mathcal{U}(x)=\frac{e^{\frac{cp\lambda}{2}x}}{(1+e^{p\lambda
x})^{\frac{a+b+1}{2}}}~ _2F_1\left(a,b,c,\frac{e^{p\lambda
x}}{1+e^{p\lambda x}}\right)\label{e12}
\end{equation}
Another linearly independent solution of (\ref{e18}) at the same
factorization energy can be written as \cite{r16}
$${\mathcal{U}}(x)=\frac{e^{p\lambda \left(1-\frac{c}{2}\right)
x}}{(1+e^{p\lambda x})^{\frac{a+b-2c+3}{2}}}~
 _2F_1\left(a-c+1,b-c+1,2-c,\frac{e^{p\lambda x}}{1+e^{p\lambda
 x}}\right)$$
The linear combination of above two solutions can be taken as the
most general non-normalizable solution of the equation (\ref{e18})
at the factorization energy $\mu$, and is
 \begin{equation}
 \begin{array}{ll}
 \displaystyle
 \mathcal{U}(x)=\alpha\frac{e^{\frac{cp\lambda}{2}x}}{(1+e^{p\lambda
x})^{\frac{a+b+1}{2}}}~ _2F_1\left(a,b,c,\frac{e^{p\lambda
x}}{1+e^{p\lambda
x}}\right)\\
\displaystyle ~~~~~~~~~~~~~~~~~~~~~~~~~~
+\beta\frac{e^{p\lambda\left(1-\frac{c}{2}\right)
x}}{(1+e^{p\lambda x})^{\frac{a+b-2c+3}{2}}}~
 _2F_1\left(a-c+1,b-c+1,2-c,\frac{e^{p\lambda x}}{1+e^{p\lambda
 x}}\right)\label{e86}
 \end{array}
 \end{equation}
 where $\alpha ~\mbox{and}~\beta $ are two arbitrary constants.
 Consequently the bound state solutions of the equation
 (\ref{e15}) for the potential (\ref{e88}), are obtained from equation (\ref{e86}) by putting $\alpha=1,~\beta = 0$
 and $a = -n,~b=1-a+\sigma+\delta,~c=1+\sigma$ (see 15.4.6 of ref. \cite{r16})
 \begin{equation}
 \psi_n(x)=\left(\frac{p\lambda (2n+\sigma+\delta+1) n!~\Gamma(n+\sigma+\delta+1)}
 {\Gamma(n+\sigma+1)\Gamma(n+\delta+1)}\right)^{1/2} \frac{e^{\frac{p\lambda(\sigma+1)}{2}}x}{\left(1+e^{p\lambda
 x}\right)^{\frac{\sigma+\delta+2}{2}}}\mathcal{P}_n^{(\sigma,\delta)}
 \left(\frac{1-e^{p\lambda x}}{1+e^{p\lambda x}}\right)\label{e89}
 \end{equation}
 and the energy eigenvalues are given by
 \begin{equation}
 E_n=n^2p+np(\sigma+\delta+1)+\frac{(\sigma+1)p(\delta+1)}{2}~,~~~n=0,1,2,...\label{e90}
 \end{equation}
 It should be mentioned here that for $\psi_n(x)$ to be a physically
 acceptable solution it should satisfy the following two conditions:\\
 (i) It should  be square integrable over domain of definition {\bf{D}} of $M(x)$
 and
 $\psi(x)$ i.e.,$$\int_D |\psi_n(x)|^2 dx<\infty$$\\
 (ii) The Hermiticity of the Hamiltonian (\ref{e15}) in the Hilbert
  space spanned by the eigenfunctions of the potential $V(x)$ is ensured by the following extra
 condition \cite{r17}
 $$\frac{|\psi_n(x)|^2}{\sqrt{M(x)}}\rightarrow 0$$ at the end points of the interval where $V(x)$ and $\psi_n(x)$ are
  defined. This condition imposes an additional restriction whenever the mass function $M(x)$ vanishes at any one or both
  the end points of $\mathcal {D}$. In order to satisfy this two conditions
  we have to impose a restriction $\sigma>-\frac{1}{2}~ \mbox{and}~ \delta>-\frac{1}{2}$ or equivalently
 $c>\frac{1}{2}~\mbox{and}~a+b-c+\frac{1}{2}>0.$
 \begin{acknowledgments}
One of us (RR) is grateful to the
Council of Scientific and Industrial Research (CSIR) New Delhi,
for a grant (project No. 21/0659/06/EMR-II).
\end{acknowledgments}


\begin{thebibliography}{99}
\bibitem{r22} If $T$ is a linear $n$th order differential operator
$Tu=\sum_{k=0}^n a_k D^k u$, then its adjoint is given by $T^*
u=\sum_{k=0}^n (-1)^k D^k(a_ku).$
\end{thebibliography}
\end{document}